\documentstyle[preprint,aps,epsfig]{revtex}
\tightenlines

\begin{document}
\draft
                 \def\gappeq{\mathrel{\rlap {\raise.5ex\hbox{$>$}} 
                 {\lower.5ex\hbox{$\sim$}}}}
                 \def\lappeq{\mathrel{\rlap{\raise.5ex\hbox{$<$}}
                  {\lower.5ex\hbox{$\sim$}}}}
         \def \gsim{\lower.8ex\hbox{$\sim$}\kern-.75em\raise.45ex\hbox{$>$}\;}
         \def \lsim{\lower.8ex\hbox{$\sim$}\kern-.8em\raise.45ex\hbox{$<$}\;}
                      \def\ltsima{$\; \buildrel < \over \sim \;$}
                      \def\simlt{\lower.5ex\hbox{\ltsima}}
                       \def\rtsima{$\; \buildrel > \over \sim \;$}
                      \def\simrt{\lower.5ex\hbox{\rtsima}}
          \def\gappr{\lower 3pt\hbox{$\buildrel > \over \sim\;$}}
          \def\gappl{\lower 3pt\hbox{$\buildrel < \over \sim\;$}}

\title{ Zeros, dips and signs in pp and p$\rm\bf\bar p$ 
       elastic amplitudes}

\author { Fl\'avio Pereira$^{1\star}$  \footnotetext{ ${}^*$ E-mail:
flavio@obsn.on.br}
and Erasmo Ferreira $^{2\star}$
\footnotetext{ ${}^*$ E-mail:erasmo@if.ufrj.br}  }  
\address{$^1$ Observat\'orio Nacional, CNPq, Rio de Janeiro 20921-400, RJ, Brazil
\\[0.1cm]
 $^2$Instituto de F\'{\i}sica, Universidade Federal do Rio de  
Janeiro \\
C.P. 68528, Rio de Janeiro 21945-970, RJ, Brazil\\  }

\maketitle

\begin{abstract}
  The dips observed in the differential cross sections of elastic pp 
and p$\bar{\rm p}$ scattering are studied in terms of the locations of the 
zeros of the real and imaginary parts of the  amplitude
and of the sign of real part at large  $|t|$. It is confirmed that
the differences in shapes of the dips in the pp and p$\bar{\rm p}$ systems 
are determined by a change of sign of the real tail, which seems to be 
determined by perturbative QCD contributions.
\end{abstract}

\newpage


\section{ Introduction} 

  The elastic differential cross sections of pp and p$\bar{\rm p}$ scattering at high 
energies present a strong forward peak, decreasing exponentially from $|t|=0$, and 
forming a dip in the range of values of transferred momentum between 1.3 and 1.5 
${\rm GeV^2}$.  For larger values of $|t|$ there is a flatter tail that, for beam energies 
above 400 GeV ($\sqrt{s}=27.5~{\rm GeV}$), seems to be independent of the energy.  
  
  Chou and Yang \cite{ChY} studied in an eikonal framework model the hadron-hadron scattering 
at ultra-high energies, predicting the existence of many dips. Using an impact parameter 
representation for the scattering amplitude,  with a $t$ dependence inspired in the proton 
electromagnetic form factor, Bourrely, Soffer and Wu \cite{WU} were able to reproduce the 
general features of the ISR experiments. Fran\c ca and Hama \cite{FH} 
described pp 
scattering under the assumption of a pure imaginary amplitude with two zeros 
parametrized as a sum of exponentials.  With a similar parametrization and including a real 
part in the amplitude, Carvalho and Menon \cite{CM} gave a more detailed representation for 
pp differential cross sections at the ISR energies. Geometrical models  in the eikonal 
approximation, including the pomeron exchange, were studied by Covolan and collaborators 
\cite{Cov}.  Extensive descriptions of  the phenomenology 
of the elastic hadron scattering can be seen in the review articles of 
Bloch and Cahn, and of Jenkovszky \cite{BCJ}.    
 
  In a somewhat detailed dynamical scheme,  Donnachie and Landshoff \cite{DLa} 
described the structure of the dip in high energy scattering through the interference of 
single-pomeron, double-pomeron and three-gluon exchanges, predicting that the dip 
in p$\bar{\rm p}$ would be less pronounced than in pp scattering.

\bigskip

  Table \ref{data}  shows the  available data on total cross section, slope parameter $B$ and ratio 
$\bar\rho$ of the forward real to imaginary parts of the amplitudes in pp and p$\bar{\rm p}$
scattering, which come from Fermilab [\cite{RRDAT1}a,f], CERN-ISR [\cite{RRDAT1}b,c,d] and 
CERN-SPS [\cite{RRDAT1}e].  Most measured differential cross sections \cite{RRDAT1,RRDAT2} 
are limited to $|t|<10~{\rm GeV}^2$, while large angle data are available only at  
$\sqrt{s}=27$ GeV \cite{Fais}, presenting a $|t|$ dependence approximately of the form 
$|t|^{-8}$ \cite{DLb}, the magnitude of $d\sigma^{e\ell}/dt$ at a given 
large $|t|$ 
being nearly energy independent \cite{Con}. In measurements of the  
differential cross section at $\sqrt{s}=19$ GeV for values of $|t|$ in the 
range 5 - 12 ${\rm GeV}^2$  \cite{Fais} the  data points converge to those 
of $\sqrt{s}=27$ GeV for  $|t| \approx 11~{\rm GeV}^2$.  In order to 
maintain the universality of the tail we have adopted in our parametrization  
for all energies between 19 and 63 GeV   the same 27 GeV values for 
large $|t|$.

 In this work we explain the detailed shapes of the dips appearing in $d\sigma^{e\ell}/dt$ 
in terms of the locations of the only one zero of the imaginary part and of 
the two zeros (in the pp case) 
of the real part.  The amplitudes are obtained from the parametrization \cite{PF} of the 
amplitudes that was inspired in the Model of the Stochastic Vaccum 
(MSV) \cite{DFK}, with additional freedom in parameters. 
The differential cross section at large $|t|$ is described through a term in the real part of the amplitude, 
and a change of sign of this term leads from the pp to the p$\bar{\rm p}$ system, the effect 
being illustrated by the analysis of the 53 GeV data.
 
  The paper is organized as follows.  In Sec. 2 we recall the parametrization 
used to describe the total and differential cross section data . In Sec. 3 
the behavior of the 
dips of the differential cross section is described in terms of the locations of the zeros 
of the real and imaginary amplitudes. Sec. 4 discusses the large $|t|$ behavior of the 
differential cross section of pp and p$\bar{\rm p}$ systems, and finally in Sec. 5 we present 
comments and conclusions. 


\section { Parametrization of the amplitudes}
 
 We use \cite{PF} the dimensionless scattering amplitude 
\begin{equation} 
T(s,t)=4\sqrt{\pi}s[i{\cal I}(t)+{\cal R}(t)]~,
\label{E36}
\end{equation}
with the elastic differential cross section given by
\begin{equation}
{d\sigma^{e\ell}\over dt}~=
          ~\frac{1}{16\pi s^2}\vert T(s,t)\vert^2 ~.
\label{dcs}
\end{equation}
The imaginary and real  parts of the amplitude are respectively parametrized in the forms
\begin{equation}
{\cal I}(t)\equiv \alpha_1{\rm e}^{-\beta_1|t|}+\alpha_2{\rm e}^{-\beta_2|t|}+
\lambda 2\rho{\rm e}^{\rho\gamma}A_\gamma(t) ~ 
\label{cd1}
\end{equation}
and 
\begin{equation}
{\cal R}(t)\equiv \alpha^{\prime}_1{\rm e}^{-\beta^{\prime}_1|t|}+\lambda^\prime
2\rho{\rm e}^{\rho\gamma^\prime}A_{\gamma^\prime}(t) ~ ,
\label{cd2}
\end{equation}
where
\begin{equation}
A_\gamma(t)\equiv {{\rm e}^{-\gamma\sqrt{\rho^2+a^2|t|}}
            \over{\sqrt{\rho^2+a^2|t|}}}-
           {\rm e}^{\rho\gamma}{e^{-\gamma\sqrt{4\rho^2+a^2|t|}}
               \over{\sqrt{4\rho^2+a^2|t|}}} ~,
\label{cd3}
\end{equation}
with $\rho= 3\pi/8$, and we have grouped the factors $2\rho e^{\rho\gamma}A_\gamma(t)$ 
in order to have $2\rho e^{\rho\gamma}A_\gamma(0)=1$.  

  These apparently complicated forms were inspired in the MSV parametrization \cite{PF} 
for the imaginary part of the scattering amplitude. The use of $A_\gamma(t)$ in the real 
part, has a more convenient structure, compared to simple exponentials, to fill the dip 
left by the zero of the imaginary part. On  the other hand, the simple exponential term 
$\alpha^\prime_1~{\rm exp}(-\beta^\prime_1\vert t\vert)$ was included in the real part 
specifically to describe the large $\vert t\vert$ ($5<\vert t\vert<15~\rm{GeV}^2$) data 
\cite{Fais} at 27 GeV.  This term is universal (energy independent) for all pp and has 
opposite sign for p$\bar{\rm p}$ ISR data.  The exponential is made numerically equivalent 
to $|t|^{-8}$ in the $|t|$ range of interest, and was used to avoid the singularity at the 
origin. This term is not used in the description of the 546 and 1800 GeV data, where large 
$|t|$ values have not been measured.  

  At $t=0$, the optical theorem and the value of $\bar\rho={\cal R}(s,0)/{\cal I}(s,0)$ 
fix the constraints
\begin{equation}
{\cal I}(0)=\alpha_1+\alpha_2+\lambda={\sigma^T\over 4\sqrt{\pi}}~,
\label{cd4}
\end{equation}
and 
\begin{equation}
{\cal R}(0)=\lambda^\prime+\alpha^{\prime}_1 =
               \bar{\rho}(\alpha_1+\alpha_2+\lambda).
\label{cd5}
\end{equation}

The values of the parameters are given in Table \ref{param}, which is an update of our 
previous determination \cite{PF}, and now includes the 
$\sqrt{s}=19.4~{\rm GeV}$ data.  The 
smoothness of the energy dependence of all parameters must be remarked.
  In comparison to the previous values, the parameters $\gamma$ and $\gamma^\prime$ have 
been slightly modified to improve the description of the data in the region of the dips of 
the pp differential cross sections as shown in Fig. \ref{dips} for 
$\alpha_1^\prime>0$ (solid lines). 
The case $\alpha_1^\prime<0$ (dashed lines), which applies to p$\bar{\rm p}$  scattering, 
is discussed latter. 

\bigskip
 


\section { Zeros of the amplitudes and dips in \lowercase{pp} scattering}

  The characteristic shape of the dip region of the differential cross sections 
has been described by Donnachie and Landshoff \cite{DLa} in 
terms of various mechanisms of pomeron and gluon exchanges.  The low $|t|$ region is 
described by the single-pomeron exchange (P), which is dominant in this region and 
gives the value to the slope $B$.  In order to describe the data, the double-pomeron 
exchange (PP) was introduced with the magnitude of its imaginary part chosen so as to 
cancel that of the P mechanism in the region where a dip is to be formed.  To yield a 
dip, the real part of the P term is partially cancelled by the three-gluon exchange 
(ggg), which is dominant for large $|t|$, has opposite signs for pp and p$\bar{\rm p}$ 
systems, and led to the prediction that at high energies the dips would be less pronounced 
in p$\bar{\rm p}$ scattering . In addition to these contributions there is the gg 
($\rho,~\omega,~f,~A_2$) exchange which is important only at very small $|t|$. We remark that 
the parametrization given by Eqs. (\ref{cd1}) and (\ref{cd2}) 
incorporates all information about the dynamical mechanisms of pp scattering
and corresponds to the sum of all terms discussed by Donnachie and 
Landshoff \cite{DLa}.  

  According to phenomenological descriptions, the structure of 
forward pp scattering 
is determined mainly by the imaginary part of the amplitude which decreases exponentially 
from $|t|=0$ and vanishes with a zero located in the interval $|t|=1.3~-~1.5~{\rm GeV}^2$.  
This zero is partially filled by the real part of the amplitude which is positive at $|t|=0$ 
and also decreases nearly exponentially, becoming negative as $|t|$ reaches 
$0.2~-~0.3~{\rm GeV}^2$.  

  Fig. \ref{exp23p} shows the separate contributions $|{\rm Im}~T(s,t)|^2$ and 
$|{\rm Re}~T(s,t)|^2$ to 
the differential cross section at $\sqrt{s}=23~{\rm GeV}$. 
 The real part in pp scattering (dashed line) has two zeros, 
becoming negative in the region between them.  The  zero of 
${\rm Im}~ T(s,t)$ (solid line) is situated between the two zeros of 
${\rm Re}~ T(s,t)$. 

\bigskip

 
  This description is valid for all energies of pp scattering from 19 to 63 GeV.  
  The zeros of the real and imaginary parts of the pp amplitude as a function of the 
energy are shown in Fig. \ref{zamp}.  The position of the zeros of 
${\rm Im}~ T(s,t)$ and the 
first zeros of ${\rm Re}~ T(s,t)$ decrease monotonically with the energy, while that 
of the second zero of ${\rm Re}~ T(s,t)$ oscillates, being lowest at about 30 GeV, as 
shown in Fig. \ref{zamp}. The position of the dips in the differential cross sections are 
always situated between the zeros of ${\rm Im}~ T(s,t)$ and the second zeros of 
${\rm Re}~T(s,t)$.  The dips are close to the zeros of ${\rm Im}~T(s,t)$, but the 
shapes in the dip region are strongly influenced by the distance between the zeros of 
${\rm Im}~T(s,t)$ and ${\rm Re}~T(s,t)$.  At about 30 GeV they are closest, and 
correspondingly the dips are remarkably pronounced (narrow and deep), as can be seen in 
Fig. \ref{dips} 
(solid lines), where the changes in the form and position of the dips in the interval 
19 - 63 GeV are exhibited in an extended scale.

\bigskip


  A similar behavior is also shown by the magnitudes of the differential cross sections at 
the dip. Fig. \ref{dsigm} shows the minimum values of $d\sigma^{e\ell}/dt$ as a function of 
$\sqrt{s}$, the lowest dip occurring at 30 GeV. The values at 546 and 1800 GeV must be 
understood only as estimates, while at 630 GeV [\cite{RRDAT1}e] the dip is 
clearly seen in the data.

\bigskip


\section {Signs of the large angle {\lowercase{pp}} and 
              {\lowercase {p$\rm\bf\bar p$ }}  amplitudes}
                     
   Donnachie and Landshoff observed that the quasi-$|t|^{-8}$ dependence of 
$d\sigma^{e\ell}/dt$ at large $|t|$ observed experimentally can be described by a 
three-gluon exchange  mechanism, which is energy independent \cite{DLb}.  Other terms 
contributing  to the amplitude are the two-gluon-one-pomeron exchanges (ggP) and three 
pomeron exchanges (PPP). The $A_{ggP}(s,t)$ and $A_{PPP}(s,t)$ terms have very little effect 
for the amplitude $A(s,t)=A_{ggg}(s,t)+A_{ggP}(s,t)+A_{PPP}(s,t)$, the dominant term at large 
$|t|$ being $A_{ggg}(s,t)$. This term has the form
\begin{equation}
A_{ggg}(s,t)~=-\frac{N}{|t|}\frac{5}{54}\biggl[4\pi\alpha_s(|\hat t|)
\frac{1}{m^2(|\hat t|)+|\hat t|}\biggr]^3~,
\label{DL1}
\end{equation}
where $m^2(|\hat t|)$ is an effective gluon mass, $\alpha_s(|\hat t|)$ is the running 
coupling constant, $\hat t\simeq t/9$, $\Lambda=200~{\rm MeV}$ and $m_0=340~{\rm MeV}$   
and for $\hat t\gg 1$ yields the $|t|^{-8}$ dependence of the 
large $|t|$ differential cross section.  The normalization 
factor $N$ is negative and determined by the proton wave function.

  To avoid the non regular behavior at small $|t|$, we use in the real 
amplitude an 
exponential form $\alpha^\prime_1\exp (-\beta^\prime_1|t|)$ rather than 
a negative power $|t|^{-8}$  to reproduce the tail.  This term describes 
the large $|t|$ behavior  at 53 GeV where both 
pp and p$\bar{\rm p}$ differential cross sections have been measured beyond the dip region.  
To show the effect of the sign of the first term in Eq.(\ref{cd2}), in 
Fig. \ref{dips}  we disregard 
the differences in the experimental values of $\sigma^T$, $B$ and $\bar\rho$ for pp and 
p$\bar{\rm p}$ systems and also keep the numerical values of all parameters of the real and 
imaginary parts of the amplitude, only changing  the sign of $\alpha_1^\prime$ in 
Eq. (\ref{cd2}). Fig. \ref{dips} shows the effects of this change of sign in the dip region, with the 
dashed lines for $\alpha_1^\prime<0$ representing p$\bar{\rm p}$ scattering. 

\bigskip


  Fig. \ref{pap53} shows $d\sigma^{e\ell}/dt$ for pp ($\alpha_1^\prime>0$) and 
p$\bar{\rm p}$ ($\alpha_1^\prime<0$) scattering at 53 GeV, where we have used, in the 
constraint imposed through Eq. (\ref{cd5}), the realistic values of $\bar\rho$ given 
for each case in Table \ref{data}.  The p$\bar{\rm p}$  dip is less deep, 
and becomes still 
flatter as the energy increases. The explanation for this behavior in terms of 
the zeros of the amplitudes is shown in Fig. \ref{exp23p}.  The real part of the 
amplitude with a positive tail ($\alpha_1^\prime>0$)  presents two zeros, the 
zero near the origin having no influence in the dip region.  
The second zero, which  is close to the zero of the imaginary part 
(solid line) is responsible for the 
pronounced form of the dip.  Changing the sign of 
$\alpha_1^\prime$, the second zero does not exist (or it would be located 
very far away).  
The change of sign in $\alpha_1^\prime$ makes, in the region of the dip, 
the ($-$) line higher than the (+) line, flattening the dips.


  \section {Comments and conclusions}

  In the present work, starting from a parametrization \cite{PF} suggested by the 
Model of the Stochastic Vaccum \cite{DFK}, we relate the behavior of the dips of the 
elastic differential cross section to the locations of the zeros of the amplitudes in 
pp and p$\bar{\rm p}$ scattering,  and investigate the role of the sign of the large $|t|$ 
tail for these systems, in the range $\sqrt{s}=$19 - 63 GeV.  

  The pp amplitude presents one zero in the imaginary part 
and two zeros in the real part. The depth of the dips is determined by the 
proximity  between the zeros of the imaginary part and the second zeros of 
the real part, so that the dips become deeper when these zeros are closer 
to each other.  In p$\bar{\rm p}$ scattering 
the real amplitude has only the first zero, which occurs far away from the dip 
region.

  The large $|t|$ tail of the differential cross section is described by an 
exponential term included in the real part of the amplitude.  This term, which is the same 
(except for a sign) for pp and p$\bar{\rm p}$ at all energies from 19 to 63 GeV, is 
responsible for the change of shape of the dip region when we change from pp to 
p$\bar{\rm p}$ systems.  The sign (which is positive for pp scattering) 
determines flatter 
dips in p$\bar{\rm p}$ scattering.  This fact, which was pointed out by 
Donnachie and Landshoff \cite{DLa}, is here confirmed by the detailed 
description of all data.

  Our parametrization describes all data in detail and has a remarkably smooth behavior 
that allows interpolations for predictions.
\bigskip

\acknowledgements {
  The authors wish to thank M. J. Menon, A. F. Martini and P. A. Carvalho for 
information  on their work. }
%



\begin{figure}[tbh]
\caption { Fittings of elastic differential cross sections through 
Eqs.(\ref{cd1}) and (\ref{cd2}) at 19.4 GeV and ISR energies, shown in an 
extended scale only in the region of the dips of  pp scattering with 
$\alpha_1^\prime>0$ (solid lines) and 
the predictions for p$\bar{\rm p}$ scattering with $\alpha_1^\prime<0$ 
(dashed lines).  Curves and data at different energies are conveniently 
separated through multiplication by powers of 10. 
\label{dips} } 
\end{figure}

\begin{figure}[tbh]
\caption {Separate contributions of $|{\rm Im} ~ T|^2$ and $|{\rm Re}~ T|^2$ for 
$d\sigma^{e\ell}/dt$ at 23.5 GeV. The dip in the data at about 1.5 GeV$^2$ is due to a 
zero in ${\rm Im}~T$. The real part (dashed line) is positive at $|t|=0$, becomes negative 
at small $|t|$ and fills partially the zero of ${\rm Im}~T$ that produces the dip.  The 
dotted line represents the squared tail term $\alpha_1^\prime\exp(-\beta_1^\prime|t|)$ alone. 
If this term is positive, it causes a second zero, as shown by the (+) line. In the ($-$) 
line  the tail is negative and  no second zero of ${\rm Re} ~ T$ exists, lowering the depth 
of the dip.
\label{exp23p} }
\end{figure}

\begin{figure}[tbh]
\caption {Locations of the zeros of the imaginary and real parts of the 
scattering amplitude (white squares) and the positions of the dips of 
$d\sigma^{e\ell}/dt$ (black squares) as functions of energy.
\label{zamp} }
\end{figure}

\begin{figure}[tbh]
\caption {Energy dependence of the values of the elastic differential cross 
section at the bottom of the dips.
\label{dsigm} } 
\end{figure}

\begin{figure}[tbh]
\caption  {Results for the elastic differential cross section  at 
53 GeV through Eq. (\ref{cd1}) and (\ref{cd2}) with $\alpha^\prime_1>0$ (pp) 
and $\alpha^\prime_1<0$ (p$\bar{\rm p}$), and with the experimental values 
of $\bar\rho$ used in Eq.(\ref{cd5}) for each case.  The pp data and the 
solid curve are multiplied by $10^{-1}$.
\label{pap53} }
\end{figure}


\begin{center}
\begin{table}
\caption{Experimental Data    \label{data}  }
\vspace{.5 cm}
\begin{tabular}{c c c c c c c}
\hline
& $ \sqrt{s} $ & $\sigma^T $ & $B $ & $ {\rm Ref.}$&$\bar\rho$&${\rm Ref.}$\\
& $(\rm{GeV})$ & $(\rm{mb})$ & $(\rm{GeV}^{-2})$ & \cite{RRDAT1} &$ $&\cite{RRDAT1}\\
\hline
&     $19.4$&$38.97\pm0.06$&$11.74\pm 0.04$&$({\rm a})$&$0.019\pm0.016$&$ ({\rm a}) $\\
&     $23.5$&$39.65\pm0.22$&$11.80\pm 0.30$&$({\rm b})$&$0.020\pm0.050$&$({\rm c,d})$\\
&     $30.7$&$40.11\pm0.17$&$12.20\pm 0.30$&$({\rm b})$&$0.042\pm0.011$&$({\rm c,d})$\\
 pp & $44.7$&$41.79\pm0.16$&$12.80\pm 0.20$&$({\rm c})$&$0.062\pm0.011$&$({\rm c,d})$\\
&     $52.8$&$42.38\pm0.15$&$12.87\pm 0.14$&$({\rm b})$&$0.078\pm0.010$&$({\rm c,d})$\\
&     $62.5$&$43.55\pm0.31$&$13.02\pm 0.27$&$({\rm b})$&$0.095\pm0.011$&$({\rm c,d})$\\
\hline
&     $52.8$&$43.32\pm0.34$&$13.03\pm 0.52$&$(b)$&$0.106\pm0.016$&$(b,c)$\\
$\bar{\rm p}$p   & $ 541  $&$62.20\pm 1.50$&$15.52\pm 0.07$&$({\rm e})$ & $0.135\pm 0.015$ & $({\rm e})$\\
&    $ 1800$&$72.20\pm 2.70$&$16.72\pm 0.44$&$({\rm f})$ & $0.140\pm0.069$ & $({\rm f})$\\
\hline
\end{tabular}
\end{table}
\end{center}
%

\begin{center}
\begin{table}
\caption{Values  of parameters for Eqs. (3) and (4). $\alpha_1$, $\beta_1$, 
$\alpha_2$, $\beta_2$, $\lambda$ and $\lambda^\prime$ are in $\rm{GeV}^{-2}$, 
and $\gamma$, $\gamma^\prime$ are dimensionless. 
$\alpha^\prime_1=0.0031~\rm{GeV}^{-2}$ and $\beta^\prime_1=0.41~\rm{GeV}^{-2}$ 
  are the same for all ISR energies and 19.4 GeV.
 \label{param}   }
\vspace{.5 cm}
\begin{tabular}{ c c c c c c c c c c }
\hline
$\sqrt{s}$&$\alpha_1$&$\beta_1$&$\alpha_2$&$\beta_2$&$\lambda $&$\lambda^\prime$&$\gamma $&
$\gamma^\prime$&$\chi^2$\\
 & & & & & & & & &\\
\hline
$~~19.4~$&$1.7400$&$1.3398$&$2.8771$&$2.2098$&$~9.4000$&$0.2632$&$3.7019$&$3.67$&$3.53$\\
$~~23.5~$&$1.7298$&$1.4390$&$3.1091$&$2.2949$&$~9.1850$&$0.2774$&$3.8625$&$5.00$&$1.18$\\
$~~30.7~$&$1.8224$&$1.4502$&$3.1649$&$2.3299$&$~9.5467$&$0.6073$&$3.9015$&$6.70$&$1.33$\\
$~~44.7~$&$1.6699$&$1.5025$&$3.0000$&$2.1086$&$10.3650$&$0.9354$&$4.0306$&$6.65$&$4.36$\\
$~~52.8~$&$1.8500$&$1.5287$&$2.9600$&$2.1753$&$10.4630$&$1.1913$&$4.0529$&$6.80$&$3.29$\\
$~~62.5~$&$1.9272$&$1.5529$&$2.8081$&$2.1337$&$10.8201$&$1.4747$&$4.0921$&$7.35$&$1.76$\\
$~546~  $&$2.6174$&$2.1539$&$3.9061$&$2.1813$&$15.7463$&$3.0270$&$4.8190$&$7.97$&$1.23$\\
$1800~  $&$3.1036$&$2.4526$&$4.4246$&$2.4253$&$18.3315$&$3.6204$&$5.3450$&$8.60$&$2.00$\\
\hline
\end{tabular}
\end{table}
\end{center}


\begin{thebibliography} {99}
%
\bibitem{ChY}T. T. Chou and C. N. Yang, Phys. Rev. {\bf D19}, 3268 (1979).
%
\bibitem{WU}C. Bourrely, J. Soffer and T.T. Wu, Phys. Rev. {\bf D19}, 3249 
(1979); Nucl. Phys. {\bf B247}, 15 (1984); Phys. Lett. {\bf B252}, 287 (1990);
{\it ibid.} {\bf B315}, 195 (1993).
%
\bibitem{FH}H. M. Fran\c ca and Y. Hama, Phys. Rev. {\bf D19}, 3261 (1979).
%
\bibitem{CM} P.A. Carvalho and M. Menon, Phys. Rev. {\bf D56}, 7321 (1997).
%
\bibitem{Cov}R. J. M. Covolan {\it et al.}, Z. Phys. {\bf C51}, 459 (1991); 
        Nucl. Phys. B (Proc. Suppl.) {\bf 25B}, 86 (1992); Z. Phys. {\bf C58}, 109 (1993); 
        Phys. Lett. {\bf B389}, 176 (1996).
%
\bibitem{BCJ}M. M. Bloch and R. N. Cahn, Rev. Mod. Phys. {\bf 57}, 563 (1985); 
             L. L. Jenkovszky, Fortschr. Phys. {\bf 34}, 791 (1986).
%
\bibitem{DLa}A. Donnachie and P.V. Landshoff, Phys. Lett. {\bf 123B}, 345 (1983); 
  Nucl. Phys. {\bf B231}, 189 (1984). 
%
\bibitem{RRDAT1}Data on pp and $\bar {\rm p}{\rm p}$ systems. 
 (a) A. S. Carrol {\it et al.}, Phys. Rev. Lett. {\bf 33}, 928 (1974); 
     A. Schiz {\it et al.}, Phys. Rev. {\bf D24}, 26 (1981). 
     L. A. Fajardo {\it et al.}, Phys. Rev. {\bf D24}, 46 (1981);
     R. L. Cool {\it et al.}, Phys, Rev. {\bf D24}, 2821 (1981);  
     R. Rubinstein {\it et al.}, Phys. Rev. {\bf D30}, 1413 (1984);
 (b) N. Amos {\it et al.}, Nucl. Phys. {\bf B262}, 689 (1985); 
 (c) R. Castaldi and G. Sanguinetti,  Ann. Rev. Nucl. Part. Sci.  {\bf 35}, 351 (1985); 
 (d) U. Amaldi and K.R. Schubert, Nucl. Phys. {\bf B166}, 301 (1980).  
 (e) D. Bernard {\it et al.}, Phys. Lett. {\bf B171}, 142 (1986);
     C. Augier {\it et al.}, Phys. Lett. {\bf B316}, 448 (1993);
     G. Fidecaro {\it et al.}, Phys. Lett. {\bf 105B}, 309 (1981);  
 (f) N. Amos {\it et al.}, Phys. Lett. {\bf B243}, 158 (1990); 
     N. Amos {\it et al.}, {\it ibid.}{\bf B247}, 127 (1990);
        Phys. Rev. Lett. {\bf 68}, 2433 (1992). 
%
\bibitem{RRDAT2} E. Nagy {\it et al.}, Nucl. Phys. {\bf B150}, 221 (1979); 
 K.R. Schubert, {\it Tables on Nucleon-Nucleon Scattering}, in
 Landolt-B\"ornstein {\it Numerical data and Functional Relationships
 in Science and Technology}, New Series, vol.I/9a (1979);
 A. Breakstone {\it et al.}, Nucl. Phys. {\bf B248}, 253 (1984); 
    Phys. Rev. Lett. {\bf 54}, 2180 (1985).
%
\bibitem{Fais} W. Faissler {\it et al.}, Phys. Rev. {\bf D23}, 33 (1981).
%
\bibitem{DLb}A. Donnachie and P.V. Landshoff, Z. Phys. {\bf C2}, 55 (1979);
  Phys. Lett. {\bf B387}, 637 (1996). 
%
\bibitem{Con} S. Conetti {\it et al.}, Phys. Rev. Lett. {\bf 41}, 924 (1978).
%
\bibitem{PF}Pereira, F. and E. Ferreira, Phys. Rev. {\bf D58}, 014008-1 (1999).
%
\bibitem{DFK}H.G. Dosch, E. Ferreira and A. Kr\"amer,  Phys. Lett.
 {\bf B289}, 153 (1992); {\it ibid.} {\bf B318}, 197 (1993);
 Phys. Rev. {\bf D50}, 1992 (1994).
%
\end{thebibliography}
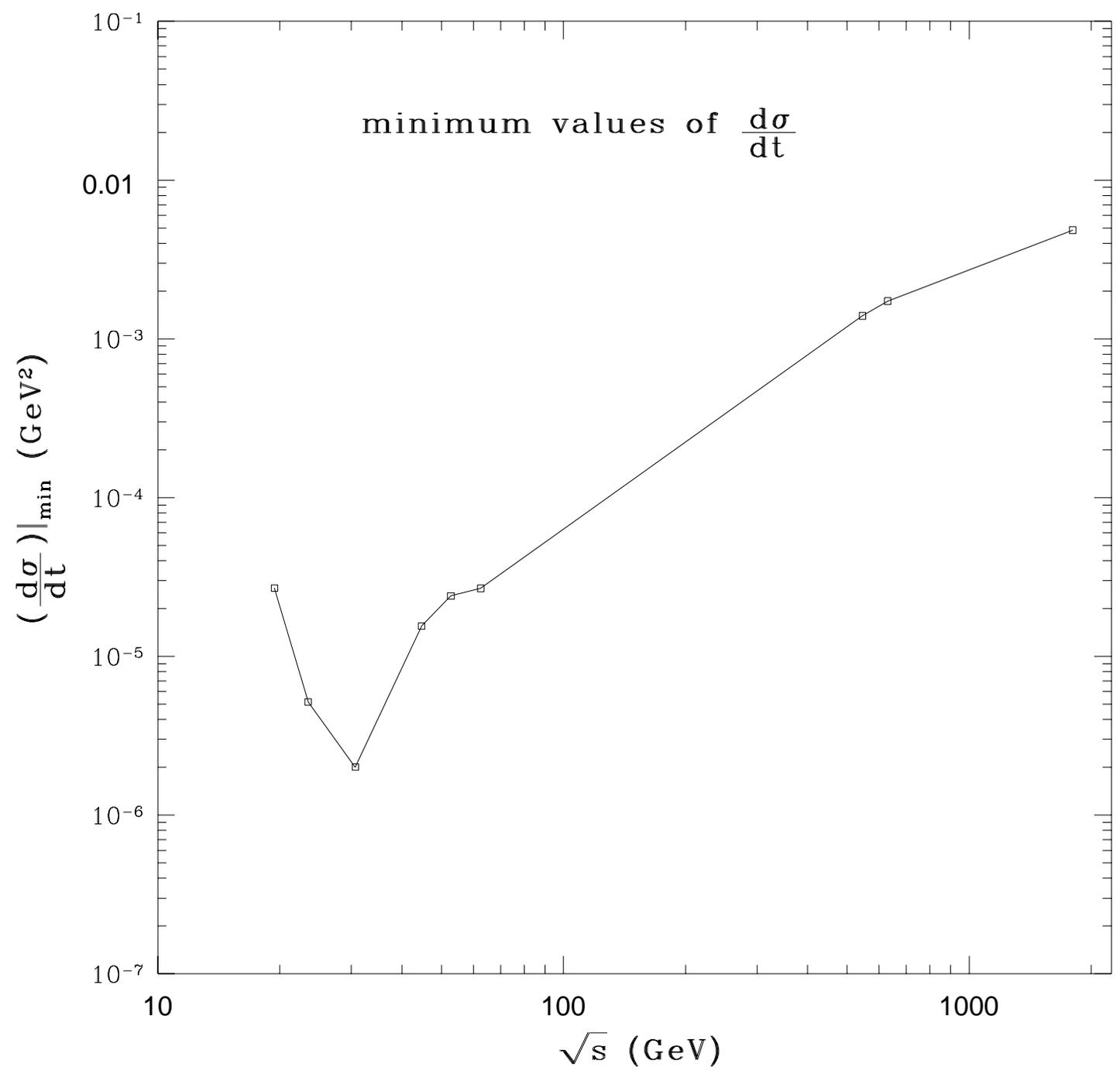
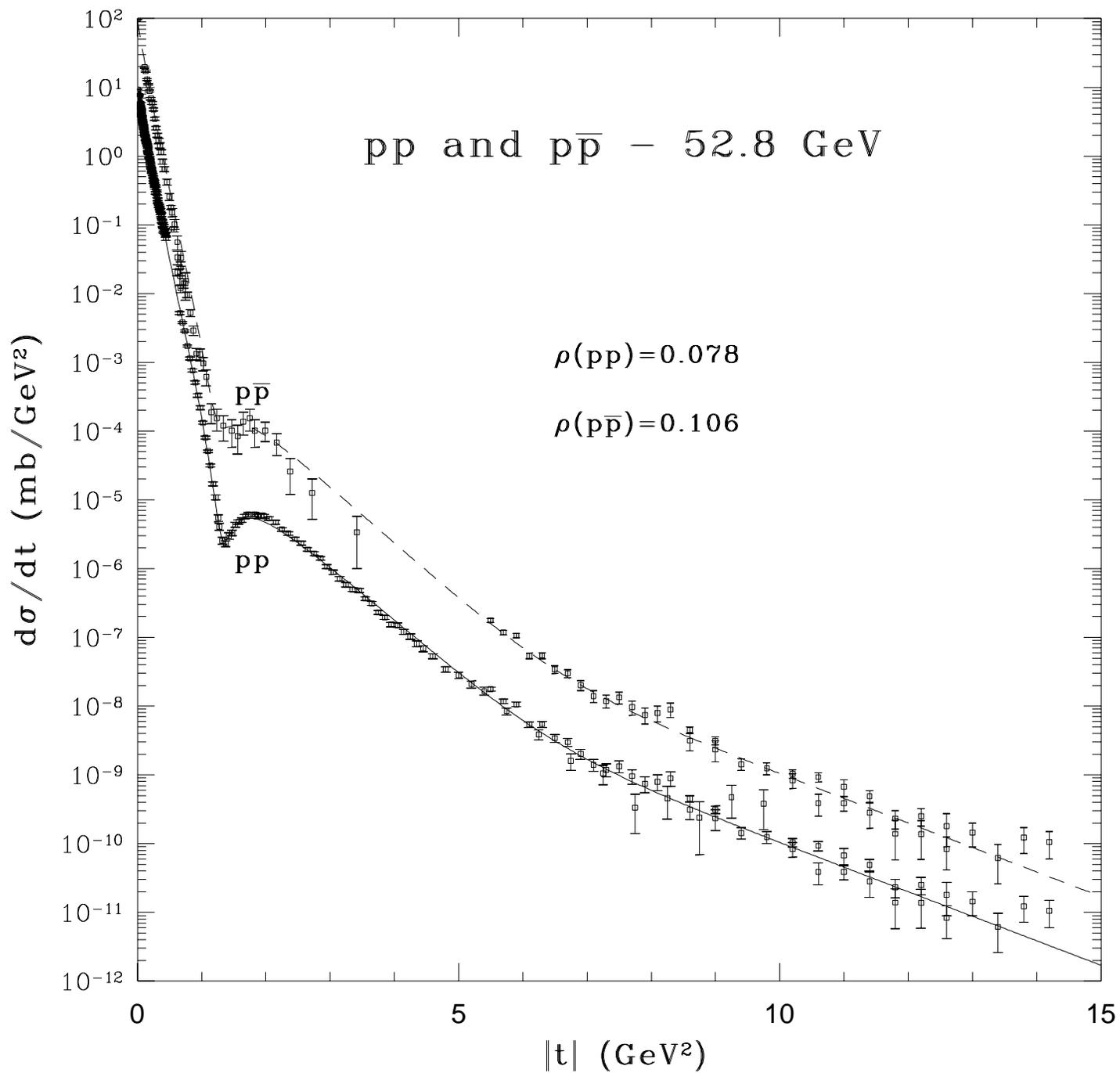
\end{document}